# Wind-induced Natural Gamma Radiation

A. Chilingarian, B. Sargsyan, D. Aslanyan, L. Kozliner

A. Alikhanyan National Lab (Yerevan Physics Institute), Yerevan AM0036, Armenia

**Abstract**

During the extreme winter storms of 2024-2025 at Aragats, natural gamma radiation (NGR) increased by more than 1000%, with fluence reaching $2\times10^7$ gammas/cm² over 10 hours and a corresponding dose of 3.26 mSv, 120 times higher than normal background radiation for the same period. This unprecedented radiation surge was detected during dry, electrified snowstorms, exceeding levels explainable by known atmospheric mechanisms, necessitating a significant reassessment of gamma-ray sources in winter storm conditions.

These results suggest similar radiation surges may occur in high-altitude and polar regions (Arctic and Antarctic), where strong winds and prolonged snowstorms are common. Understanding radiation surge conditions is essential for refining atmospheric models, improving radiation monitoring, and assessing environmental and climatic impacts in extreme weather conditions.

**Key points**

• We observe an unprecedented enhancement of gamma radiation during snowstorms on Aragats.

• We propose a *Wind-Induced Gamma Radiation Enhancement* mechanism that extends beyond RREA-driven and radon progeny models.

• Gamma radiation increased by more than 1000% during strong winds.

• **New revealed phenomenon** influence ionization rates, atmospheric conductivity, and climatic conditions.

**1.Introduction**

Natural gamma radiation (NGR) plays a significant role in the Earth's environment, affecting various areas such as radiation protection, geological exploration, atmospheric chemistry, and astrophysics. Ongoing scientific research aims to understand and monitor NGR sources and levels for human safety and scientific investigation. In this context, we examine a new atmospheric source of NGR that arises during very strong winds.

Electric fields generated in the atmosphere during thunderstorms effectively accelerate electrons and initiate electron-photon avalanches when the field strength exceeds the critical value specific to air density (Gurevich et al., 1992). When the electric field is sufficiently strong, electron and gamma-ray fluxes are detected on the ground as thunderstorm ground enhancements (TGEs, Chilingarian et al., 2010; 2011). TGEs can last a few seconds to several tens of minutes, energy spectra can extend up to 60 MeV, and fluence can reach tens of particles per cm² (Chilingarian et al., 2023; 2024a; 2024b). If the electric field falls below the critical value, a minor enhancement of gamma-ray flux may occur due to the energy gain from the electric field. This leads to an alteration of the electron energy spectrum (MOS effect, Chilingarian et al., 2012a), which increases the likelihood of bremsstrahlung.

Suppose an accelerating electric field terminates above the surface (>150 m). In that case, electrons will not reach the surface due to ionization losses, and only gamma rays contribute to the TGE flux, forming gamma glows (Wada et al., 2021). If the gamma flux is enormously large, neutrons produced in photonuclear reactions can be detected by neutron monitors (Chilingarian et al., 2012b; Kishvardai et al., 2024) and SEVAN detectors (Chilingarian et al., 2018) located at mountaintops.
During a thunderstorm, radon progeny emanating from basalts is lifted into the atmosphere by the near-surface electric field (NSEF) and significantly contributes to natural gamma radiation (NGR) from $^{214}$Pb, $^{214}$Bi, and other isotopes, all below 3 MeV (Radon circulation effect, Chilingarian et al. 2020).

Although TGEs predominantly occur in spring and autumn, they are rare in summer and especially in winter (Chilingarian et al., 2021; 2024a). However, in December 2024, during subzero temperatures (< -10°C) on Mount Aragats, unusual flux enhancements were observed, correlated with strong winds. In the highlands and large northern regions, horizontal wind flows traverse extensive areas covered with dry snow. These winds strike outdoor particle detectors and fill the buildings housing them with air enriched in radon progeny and electrified snow. Dry snow, lifted and charged by strong winter winds, may form compact, electrified aerosol structures with elevated radon concentrations and their radioactive progeny. Blowing into detector housings and penetrating laboratory spaces, these radioactive clouds can sustain enhanced gamma-ray fluxes for many hours.

As a result, the gamma radiation from radon progeny significantly elevates particle detector count rates. Moreover, repeated mechanical interactions between dry, charged snow particles and metallic components of atmospheric electricity sensors (e.g., field mills and lightning antennas) induce spurious discharges via the triboelectric effect. These discharges generate false spikes in the near-surface electric field (NSEF) and may trigger erroneous lightning locations, mimicking natural electrostatic phenomena and confounding atmospheric diagnostics. Thus, during strong snowstorms, part of the recorded electric field disturbances and lightning activity may be artifacts of triboelectric interactions, not genuine atmospheric discharges.

We refer to the observed winter phenomena as wind-induced gamma radiation enhancements (WiGERs), which differ from thunderstorm-driven TGEs. We analyze WiGERs in detail, focusing on the modes of particle flux enhancement linked to strong winds and dry snow. These findings introduce a new powerful source of atmospheric natural gamma radiation.

**2.Instrumentation**

In (Chilingarian et al., 2024c) we present the detailed description of Experimental facilities on Aragats. The Aragats Solar Neutron Telescope (ASNT), located in the MAKET experimental hall of 20,000 m³ volume, measures the flux of electrons and gamma rays in the energy range of 10–100 MeV. The Aragats Neutron Monitor (ArNM), type 18HM64, and the SEVAN detectors are in the same hall. A network of three STAND1 detectors (three stacked scintillators with a thickness of 1 cm and an area of 1 m$^2$, and one stand-alone with a thickness of 3 cm) is located at Aragats station premises, covering a ≈50000 m² area. DAQ electronics based on the National Instruments MyRIO board (see details in Chilingarian et al., 2024d) continuously measure and send to servers a 50-ms time series.

In the SKL experimental hall with a volume of 4,000 m³, the STAND3 stacked detector (Chilingarian and Hovsepyan, 2023), which incorporates four 3 cm thick and 1 m² area plastic scintillators, measures vertical electron fluxes with energy thresholds of 20, 30, and 40 MeV by identifying coincidences across 4 layers of the detector. The energy releases of gamma rays and electrons are measured using the SEVAN-light spectrometer, which consists of a 20 cm thick and 0.25 m² spectrometric scintillator and a 1 cm thick, 1 m² area "veto" scintillator positioned above it. The SEVAN-light detector records energy-release histograms continuously each minute. The efficiency of the upper 1 cm-thick veto scintillator in registering charged particles exceeds 95%, while its efficiency for neutral particles is around 2%. Thus, in an '11" coincidence, the signal in both scintillators indicates passage of charged particles, while a '01" coincidence selects the neutral ones. Therefore, using a SEVAN-light detector, we can disentangle neutral and charged fluxes and measure the energy releases of each separately. The SEVAN-light detector is located on the first floor of the SKL hall.

The Logarithmic amplitude-to-digit converter (LADC), used in SEVAN light electronics, allows for the acquisition of spectra ranging from 0.3 to 50 MeV at the cost of low energy resolution in the lowest energy band (≈60%). For recovering the energy spectra from energy release histograms, the detector response function was calculated using GEANT 4 by transporting particles through the building and detector media, considering various sources of randomness and uncertainty in the measurement process. The response matrix will account for the smearing effects due to the finite resolution of the detector and the asymmetry in the bin-to-bin migration due to very steep cosmic ray spectra. The energy spectra recovery procedure is detailed in (Chilingarian et al., 2024e). As AEF is vertical, we use the inverse matrix obtained by vertical transport of TGE electrons and gamma rays from the roof of the building through the veto scintillator down to the bottom of the 20 cm thick spectroscopic scintillator. However, when the radioactive cloud blows into the building, the gamma rays enter the scintillator from all sides without interacting with additional material. Thus, new simulations and a new inverse matrix for recovering energy spectra were performed. The count rate of the spectrometer significantly increased due to an enhanced detection surface (0.9 m² compared to 0.25 m² for vertical TGE particle transport) and a larger path in the scintillator for half of the incidences (0.5 m instead of 0.2 m for vertical transport).

An additional SEVAN-type detector is installed in a 25 m³ hut on four 5-meter-long pipes near the SKL hall (see inset to Fig.3) . This hut, known as 'Cuckoo's Nest,' is fully exposed to the wind. The 1-second time series of count rates is recorded using analog discriminator inputs for the photomultiplier signals. Light from particles entering the sensitive volume of scintillators during ≈1 μs is summarized to produce a PM impulse fed to the comparator (Rylander et al., 2004).

A network of commercially available field mills (Model EFM-100, BOLTEK1, 2025) continuously monitors the near-surface electric field (NSEF). The EFM-100's sensitivity range for lightning location is 33 km, and the instrument's response time is 100 ms. Changes in the electrostatic field are recorded at a sampling interval of 50 ms.

The lightning activity from 30 km to 480 km is monitored by Boltek's Storm Tracker (lightning detection system, BOLTEK2, 2025). Storm tracker defines four types of lightning (CG-, CG+ cloud-to-ground negative and positive, IC -, IC+ intracloud positive and negative) in radii up to 480 km around the location of its antenna. On 14 December, multiple discharges in Boltek instruments caused by dry, electrified snow encounters resulted in a whole area around the station, spanning a hundred kilometers, being detected by lightning detections. However, the worldwide lightning location network (WWLLN), one of the nodes located in Yerevan, didn't register any lightning discharges from December 14 to 28 within a 200km radius.

Davis Instruments' Vantage Pro2 Plus automatic weather stations monitor meteorological conditions, including a wind speed sensor, rain collector, atmospheric pressure sensor, temperature and humidity sensors, anemometer, solar radiation sensor, and UV sensor. When wind fills buildings with air enriched by radon progeny, gamma radiation enters detectors not only from the top (as with RREA avalanches accelerated in the vertically oriented atmospheric electric field but also from all sides of the detectors, reaching 3.6 m² for ASNT and 0.9 m² for the SEVAN light.

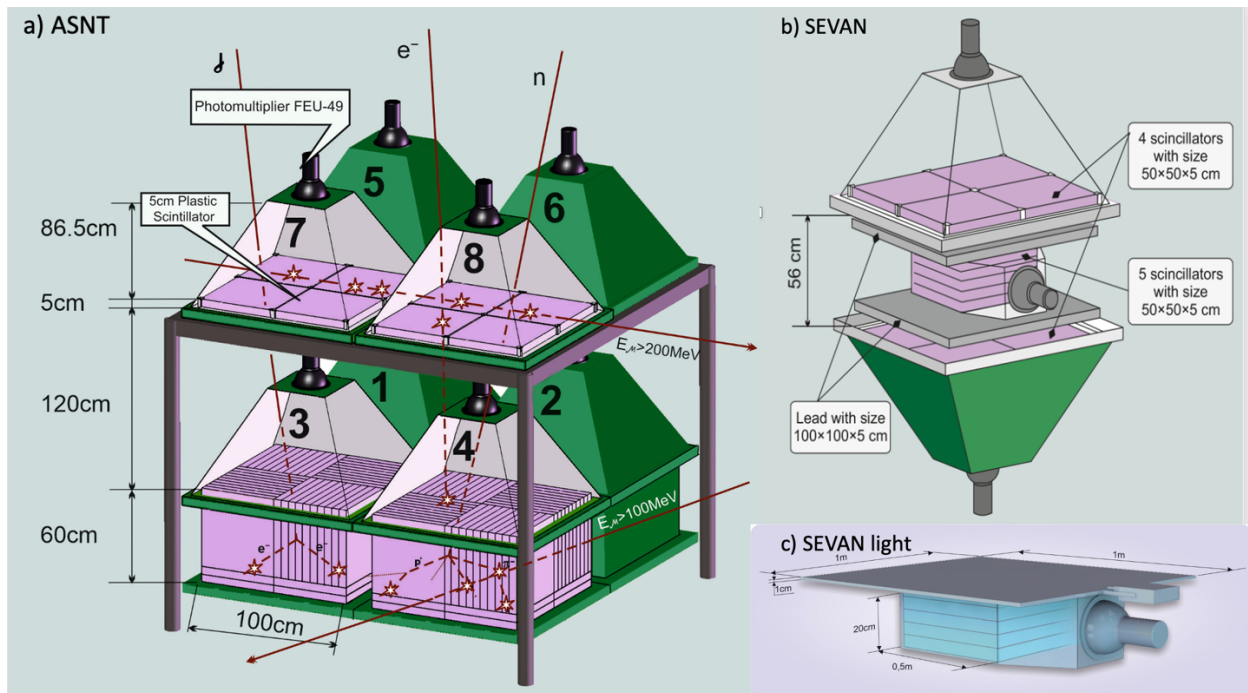

**Figure 1. Particle detectors of the Aragats research station (3200 m asl)**

### 3. Winter flux enhancements

In Winter, the atmospheric electric field usually does not reach the critical strength to unleash electron-photon avalanches. However, due to the effect of Radon circulation (Chilingarian et al., 2020) during thunderstorms, the electric field lifted charged aerosols with attached Radon progeny into the atmosphere. Gamma radiation from Radon progeny caused prolonged peaks in the time series of particle detector count rates (Chilingarian et al., 2021). Moderate TGE fluxes occasionally occur in winter, induced by electrified clouds occasionally appearing above the research station. In Chilingarian et al. (2024b), we reported TGE on February 6, 2023, in the depths of winter at -10°C (a day after the devastating earthquake in Turkey).

While analyzing winter flux enhancements, we identified an instrumental artifact caused by the electrification of the electric field mill (EFM-100) due to dry electrified snow impacting the sensor's metallic plates (triboelectric effect). Snow particles become highly charged in dry, windy, and cold conditions. During strong winds, when snow makes intense contact with the sensor's metallic plates, they discharge upon impact, altering the electric field readings. This instrumental effect induces prolonged false signals in electric field measurements. The discharges within the device can mimic natural atmospheric electrical activity in the time series of NSEF.

According to the Armenian Meteorological Service, the strongest winds of 2024 occurred between December 14 and 16, reaching speeds of up to 30 m/s. On Aragats, winds reached 20 m/s, while the outside temperature ranged from -10 to -20 degrees Celsius. Simultaneously, we observe unusual patterns of NSEF (blue curve in Fig. 2). Unusual dense blue curves, which appear only during Winter's strong winds, highlight the high-frequency fluctuations in the EFM 100 readings, demonstrating continuous discharges in the device. These episodes can last up to

10 hours as the wind blows. Simultaneously, we observe a huge increase in the gamma-ray flux (yellow curve) measured by the 20 cm thick scintillator of the SEVAN-light detector at 14:24-18:42 (258 minutes, Fig. 2). There was no increase in the coincidence of 1 cm thick and 20 cm thick scintillators ("11" coincidence, black curve in Fig. 2) that selects charged particles. Thus, only the gamma ray flux was enhanced. To gain insight into gamma ray flux enhancement and determine its origin, we recover energy spectra on a minute-by-minute basis for the whole 258 minutes of flux enhancement. The maximum energy was determined in each minute from 14:24 to 18:42 (pointed by the red arrow). The time series of maximum energies is shown in the inset to Fig. 2. The principal contributor to gamma radiation from $^{222}$Rn progeny is the isotope $^{214}$Pb, with a half-life of 26.8 minutes and emitting prominent gamma lines at 295 keV and 352 keV with emission probabilities of 18.5% and 35.8%, respectively. $^{214}$Bi with a half-life of 19.9 minutes, produces a broad spectrum of gamma rays, including a strong line at 609 keV (~45%) and several higher-energy emissions: 1120 keV (~15%), 1238 keV (~5.8%), 1378 keV (~4.0%), 1764 keV (~15%), and a maximum-energy gamma ray at 2204 keV with an intensity of approximately 5%. However, most 1-minute maximum energies exceed 3 MeV, and a few reach 9 MeV. These large maximum energies of gamma rays exceeding Radon progeny energies require clarification and confirmation to be presented in the next section.

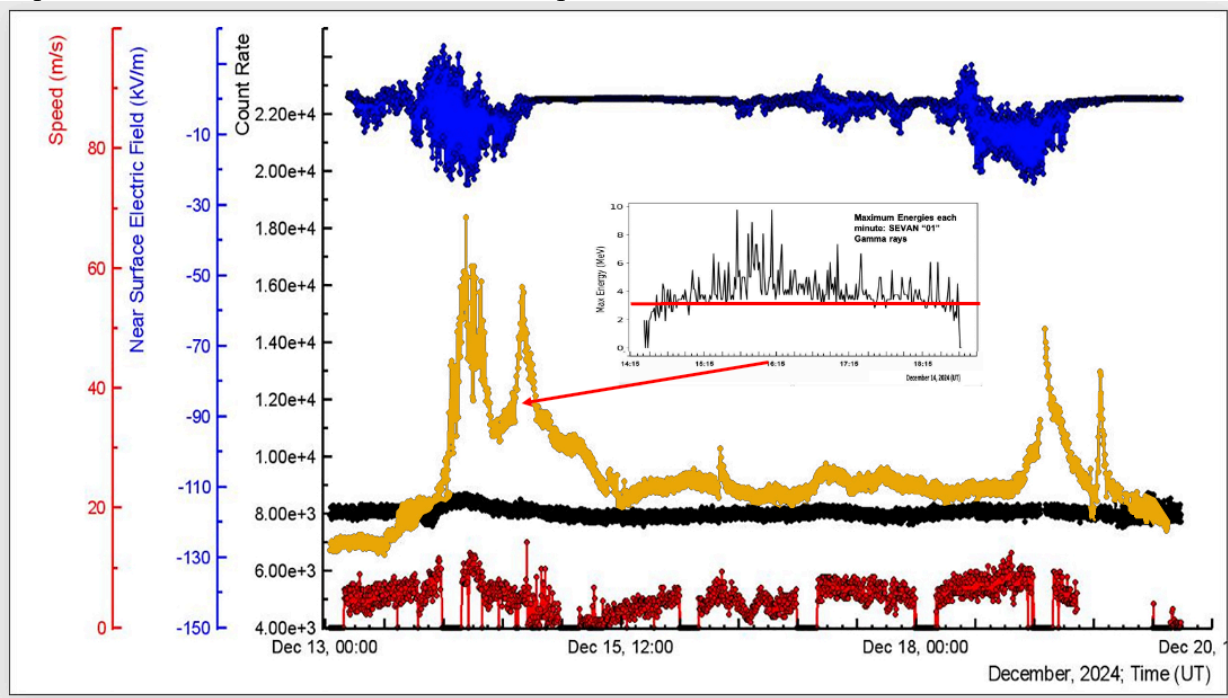

**Figure 2. One-minute time series of "11" coincidence (electrons, black) and "01" coincidence (gamma rays, brown) measured by the SEVAN-light detector; disturbances of NSEF (blue); and wind speed (red). In the inset, we display the minute-by-minute maximum energies of gamma rays; the red arrow indicates when energy spectra were selected. The red line in the inset demonstrates that the maximum energy during this period exceeds 3 MeV. Due to severe weather conditions, there are several gaps in wind measurement.**

**4. Analysis of the prolonged gamma-ray enhancements during strong winds**

From the previous section, it is apparent that there is a contradiction: the climatic conditions do not support runaway avalanches in the atmosphere that can produce particles with energies exceeding 10 MeV (no thunderstorm, no expected RREA in the atmosphere that can produce relativistic particle avalanches). Yet, the energy spectra are significantly above the radon progeny gamma radiation levels. First, we check the flux enhancements recorded by thick and thin scintillators in the Cuckoo's Nest hut (see inset to Figure 3), showing a huge enhancement at 0:0 – 9:30 the same day, 14 December 2024. Horizontally placed 1 cm thick detectors register all TGEs at Aragats (see the TGE catalog, Chilingarian et al., 2024d). However, as shown in Fig. 3, the 1 cm thick scintillator of the Cuckoo's Nest detector (lower black line), as well as the 1 cm thick scintillator of the SEVAN light detector, does not register the count rate enhancement observed by the 20 cm thick scintillators just below them (black lines in Fig. 2 and 3). The detector in Cuckoo's Nest is a SEVAN-light detector (Fig. 1c); however, it uses different DAQ electronics, which do not include an ADC, only scalers that measure particle flux each second (Rylander et al., 2004). The 1 cm-thick detectors of the STAND1 network also do not register any enhancement. Consequently, we conclude that the radiation comes not from a vertical but a horizontal direction. When strong wind blows into buildings, detectors are exposed to radon progeny, aerosols, and electrified snow from all 6 sides, with a sensitive surface of ≈ 1 m$^2$.

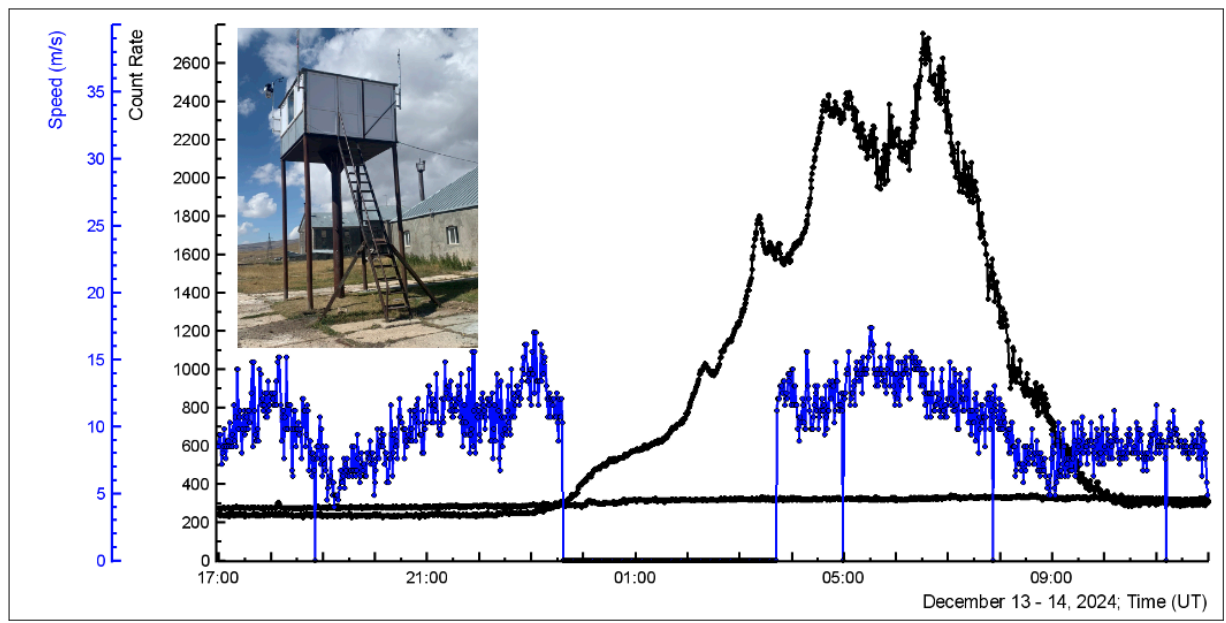

**Figure 3. 1s time series of electrons and gamma rays and 1-minute time series of wind speed. Due to severe weather conditions, there is a gap in wind measurement for ≈4 hours.**

At the maximum flux, the enhancement count rate measured by the Cuckoo's Nest detector reaches 1150% (590σ), the first yellow pic in Fig.2. After registering the record enhancement, snow entered the hut and cover electronics, which failed to register the second peak that we present in Fig. 2. During second phase of wind storm, the enhancement measured by the SEVAN light detector reaches 35% (35σ). This difference is attributed to varying exposure to strong winds carrying isotopes and electrified snow; the hut has a hole in the bottom, and wind directly blows electrified snow inside. The SEVAN light detectors are located on the first floor of the

MAKET building, covered from direct snow blows. The third spectrometer measured coherent enhancement using a 60 cm thick scintillator from the ASNT detector (Fig. 1a).

Over the 258-minute event, the total fluence was $2.58 \times 10^3$ gamma rays per square centimeter. Assuming a mean gamma-ray energy of 1 MeV and using the ICRP (ICRP, 2010) dose conversion factor of $5.7 \times 10^{-14}$ Sv per gamma, the corresponding effective dose is estimated to be 1.47 mSv. This dose significantly exceeds the natural background over the same interval and confirms measurable exposure from wind-transported radon progeny. The detector located in "cockiest nest" hut observes an intensity at least ten times greater, thus the effective dose can reach 15 mSv or more during windy days in winter.

In Figure 4, we present the differential energy spectra of randomly selected minutes (Figs 4a-4c), and the integral energy spectra average over minutes from 14:24 to 18:42 (Fig. 4d). The mean gamma-ray flux during the windstorm (14:24–18:42) was 100,000 gamma rays per minute per square meter. The observation time was 258 minutes, and the particle accepting window of the SEVAN light spectrometer was 1 microsecond. Gamma rays entered the 20 cm-thick and 0.25 $m^2$ area scintillator from six sides. The detector energy resolution was approximately 70% FWHM at 1.6 MeV, corresponding to a standard deviation $\sigma \approx 0.475$ MeV.

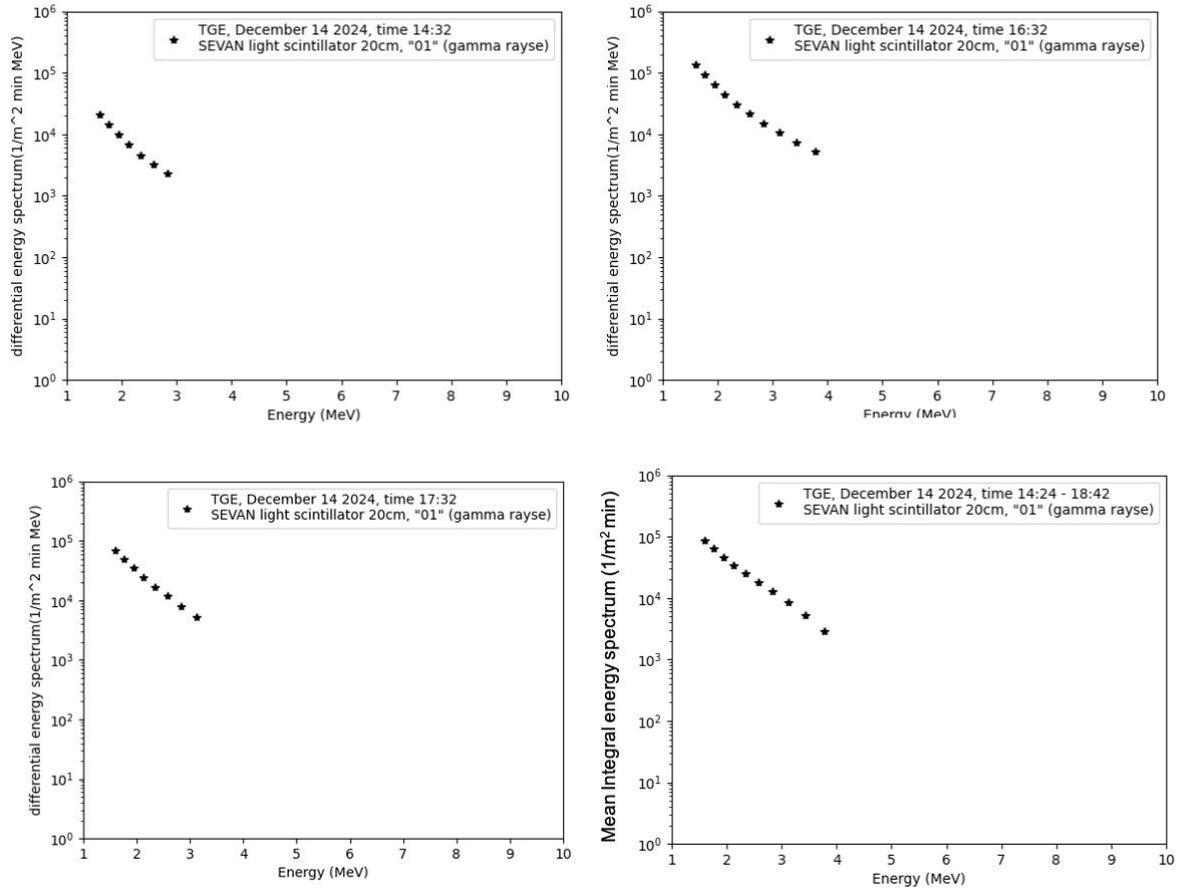

**Figure 4. a)-c) Randomly selected minute differential energy spectra recovered from the energy releases registered by the SEVAN light detector. D) Minute integral energy spectra averaged over 14:24 to 18:42.**

The average number of gamma rays arriving during a 1 microsecond window is $\lambda = 1.67 \times 10^{-3}$. The Poisson probability of two gamma rays arriving in the same window is $P(2) \approx 1.39 \times 10^{-6}$. With 60 million windows per minute, the expected number of double pile-up events is about 83 per minute. Crucially, only one such event per minute is needed to raise the maximum energy in each 1-minute histogram, so this rate is more than sufficient to explain observed maxima. The probability of triple pile-up is much smaller: $P(3) \approx 7.7 \times 10^{-10}$, corresponding to one triple pile-up every ~22 minutes. These rare triple coincidences can contribute to the high-energy tail observed in several histogram intervals, especially when accounting for detector resolution.

Due to the detector's finite resolution, pile-up energies can be significantly smeared. A double pile-up of two 1.6 MeV gamma rays produces a 3.2 MeV signal, which, when combined with a 3σ upward fluctuation (σ ≈ 0.475 MeV), yields a detected energy of 4.63 MeV. A triple pile-up (4.8 MeV) plus 3σ extends to 6.23 MeV. These effects enable low-energy gamma rays to mimic much higher energies due to pile-up and resolution smearing.

In summary, with a flux of 100,000 gamma rays per minute and a 1 μs window, more than 80 double pile-ups occur per minute, and triple pile-ups appear every 20–25 minutes. Detector resolution further boosts the apparent energy of pile-ups. Thus, pileup and poor energy resolution can explain the observed maximum energies up to 10 MeV during strong wind gamma ray flux enhancements measured by the SEVAN light spectrometer.

**Discussion and Conclusions**

During the extremely strong winds at Aragats in winter 2024-2025, natural gamma radiation levels increased by more than 1000%, during electrified snowstorms driven by intense winds. Observed radiation levels far exceed what can be explained by any known atmospheric mechanisms, indicating the presence of an unaccounted-for particle transport and radiation amplification process under winter snowstorm conditions. We propose a novel mechanism, distinct from relativistic runaway electron avalanches or radon progeny radiation models. Our model suggests that:

1. Cold air traps radon near the ground, whereas snow-covered soil reduces radon exhalation compared to warm seasons.
2. Strong wind lifts radon progeny to detector altitude, overcoming normal gravitational settling effects and dispersing radioactive aerosols over large areas.
3. Electrified snow particles act as efficient carriers, increasing radon daughter attachment rates to airborne aerosols and maintaining high radiation levels.
4. A dense radioactive cloud significantly amplifies gamma-ray flux, sustained by continuous turbulent interactions of emanated Radon progeny, dry electrified snow, and aerosols.

Unlike short-lived TGE bursts, where thunderstorm electric fields drive electron acceleration, wind storm enhancements are maintained for many hours, suggesting a continuous radon source rather than a single release event. This sustained gamma-ray enhancement indicates that strong winds are fundamental in enhancing radiation in winter storms. Thus, the Aragats observations reveal a previously overlooked mechanism of gamma-ray enhancement in high-altitude snowy environments. Similar radiation surges may occur in regions such as the Arctic and Antarctic, where katabatic winds, persistent snowstorms, and charged aerosols create conditions favorable for radioactive cloud formation.

Possible analogies can be found in other natural systems where high-energy radiation is emitted under complex plasma conditions. These systems demonstrate that significant gamma emissions can occur from charge accumulation, electrostatic field formation, and interaction with ambient radioactive materials.

The lunar surface experiences energetic emissions caused by charge accumulation on dielectric surfaces exposed to solar wind and micrometeorite impacts. The Kaguya mission recorded gamma rays from 200 keV to 13 MeV linked to interactions between solar plasma, dust, and surface electrical fields (Kobayashi et al., 2013).

The proposed mechanism could have broad implications for:

- Background radiation monitoring, where radioactive cloud could distort long-term datasets and misrepresent baseline gamma-ray flux levels.

- Aviation safety, where prolonged radiation exposure in extreme weather conditions could require revised risk assessments for surface and near-surface operations.

- Atmospheric electricity and climate models, where charged aerosols in snowstorms may influence local ionization rates, cloud microphysics, and large-scale atmospheric charge dynamics.

Understanding this process is critical for atmospheric physics, space weather studies, and environmental monitoring. Further research is necessary to quantify radioactive cloud formation across different altitudes, climates, and meteorological conditions.

## Acknowledgment

We thank the staff of the Aragats Space Environmental Center for the uninterrupted operation of all particle detectors under severe winter conditions. The authors acknowledge the support of the Science Committee of the Republic of Armenia (Research Project No. 21AG-1C012) in modernizing the technical infrastructure of high-altitude stations. A.C. thanks Suren Hovhakimyan for providing data on the strong winds in Armenia.